\documentclass[aps,prd,floatfix,nofootinbib,11pt]{revtex4-1}

\usepackage{graphicx}
\usepackage{epic}
\usepackage{eepic}
\usepackage{latexsym}

\newcommand{\df}{\dot{\phi}}
\newcommand{\zo}{\zeta_k^{(0)}}
\newcommand{\cco}{\cc_k^{(0)}}

\newcommand{\eq}[1]{(\ref{#1})}
\newcommand{\be}{\begin{equation}}
\newcommand{\ee}{\end{equation}}
\newcommand{\bea}{\begin{eqnarray}}
\newcommand{\eea}{\end{eqnarray}}

\newcommand{\hs}[1]{\hspace{#1 mm}}

\newcommand{\lf}{\left<}
\newcommand{\rg}{\right>}

\newcommand{\tm}{\tilde{M}_p}

\def\a{\alpha}
\def\b{\beta}
\def\cc{\gamma}

\def\d{\delta}

\def\e{\epsilon}

\def\f{\phi}
\def\fr{\frac}
\def\F{\Phi}
\def\vf{\varphi}

\def\l{\lambda}

\def\m{\mu}

\def\s{\sigma}

\def\th{\theta}

\def\z{\zeta}

\def\o{\omega}

\def\vf{\varphi}

\def\del{\partial}

\let\bm=\bibitem
\def\nn{\nonumber}

\begin{document}

\title{More on loops in reheating:  Non-gaussianities  and tensor power spectrum} 

\author{Nihan Kat\i rc\i}
\email[]{nihan.katirci@boun.edu.tr}
\author{Ali Kaya}
\email[]{ali.kaya@boun.edu.tr}
\author{Merve Tarman}
\email[]{merve.tarman@boun.edu.tr}
\affiliation{Bo\~{g}azi\c{c}i University, Department of Physics, \\ 34342,
Bebek, \.Istanbul, Turkey}

\begin{abstract}

We consider the single field chaotic $m^2\f^2$ inflationary model with a period of preheating, where the inflaton decays to another scalar field $\chi$ in the parametric resonance regime. In a recent work, one of us has shown that the $\chi$ modes circulating in the loops during preheating notably modify the $\lf\zeta\zeta\rg$ correlation function. We first rederive this result using a different gauge condition hence reconfirm that superhorizon $\zeta$ modes are affected by the loops in preheating. Further, we examine how $\chi$ loops give rise to non-gaussianity and affect the tensor perturbations. For that, all cubic and some higher order interactions involving two $\chi$ fields are determined and their contribution to the non-gaussianity parameter $f_{NL}$ and the tensor power spectrum are calculated at one loop. Our estimates for these corrections show that while a large amount of non-gaussianity can be produced during reheating,  the tensor power spectrum receive moderate corrections. We observe that the loop quantum effects increase with more $\chi$ fields circulating in the loops indicating that  the perturbation theory might be  broken down. These findings demonstrate that the loop corrections during reheating are significant and they must be taken into account for precision inflationary cosmology. 

\end{abstract}

\maketitle

\tableofcontents

\section{Introduction}

Single scalar field inflationary models have solid predictions for the scalar and the tensor power spectra, and  the amount of non-gaussianity produced by the interactions. These observable quantities are fixed by a few parameters like the slow-roll parameter of the potential. Moreover, in these models the quantum loop corrections to the standard inflationary predictions turn out to be quite small (see  e.g. \cite{s1,s2,s3}). As a result of this firm structure, many single field scalar models are either ruled out or severely constrained by the recent Planck data \cite{pl} (see \cite{ls} for a scan of inflationary scenarios in the light of Planck). For example, the chaotic $m^2\f^2$ model is ruled out by 95\% confidence level (provided the index is not running) by the contours in the the scalar-to-tensor ratio $r$ vs. the scalar spectral index $n_s$ data plane \cite{pl}.  

The constancy of the superhorizon curvature perturbation $\zeta$ is very crucial for the inflationary predictions to hold. This helps to determine the cosmic microwave background (CMB) fluctuations from the correlation functions evaluated at the horizon crossing time. Technically, the conservation of $\zeta$ sets an upper bound for the time integrals that appear in the in-in perturbation theory \cite{mal}. 

On the other hand, it is well known that the entropy perturbations can cause superhorizon evolution of the curvature perturbation (see e.g. \cite{mfb}). In reheating, those fields that are unimportant during the exponential expansion are exited by the inflaton decay and they start to dominate the universe. Moreover, in some models the decay can occur violently in a preheating stage \cite{reh1,reh2,reh3,reh4}. Although these entropy perturbations are not produced at cosmologically interesting scales, reheating stage ends with highly nonlinear processes  (see e.g. \cite{n1,n2,n3}). While these nonlinearities can be effectively described by fluid dynamics that only affect local quantities \cite{bsz}, some of them are known to have important consequences (see e.g. \cite{coh1,coh2}). 

Classical, and similarly quantum, nonlinearities imply that Fourier modes do not evolve independently. As a result of  this mode-mode coupling, short distance fluctuations are expected to affect the long wavelength modes. For example, a cubic interaction term in a Lagrangian would allow two modes with nearly equal large momenta to change the amplitude of a mode with small momentum. In quantum theory, there are also virtual modes circulating in the loops that affect the correlation functions. Evidently, it is crucial to determine the size of such effects. In a recent work, one of us has shown that in the chaotic $m^2\f^2$ model with a period of preheating where the inlaton $\f$ decays to another scalar $\chi$, the parametrically amplified $\chi$ modes appearing in the loops would meaningfully modify the curvature power spectrum \cite{ali1}. This is an example of the entropy perturbations affecting the superhorizon curvature variable, however not by the real physical fluctuations but because of the virtual entropy modes appearing in the loops (entropy modes are known to give power  loop infrared divergences during inflation \cite{ir1,ir2}). 

Since $\zeta$ becomes an ill defined dynamical variable during reheating, the calculations in \cite{ali1} have been carried out in the $\zeta=0$ gauge. In that case, one may first  calculate the $\chi$ loop corrections to the inflaton fluctuation power spectrum until  the coherency of the  inflaton oscillations is lost.  After that moment, the possible effects on the superhorizon evolution are expected to be averaged out and become negligible. One may then apply a gauge transformation to read the $\zeta$ power spectrum. Since in the first stage of the preheating the  background inflaton  oscillates coherently, the superhorizon modes are affected without violating causality \cite{ca}.  Note that as long as the relativistic equations are treated properly, there should not arise any issue with causality. 

In this paper, our first aim is to carry out the calculation of \cite{ali1} in the $\vf=0$ gauge, i.e. we will use $\zeta$ directly as the main dynamical variable. Because $\zeta$ is only ill defined at isolated times when the inflaton velocity vanishes, the propagator has ``spikes" and it diverges at these moments.  We smooth out these spikes by using the time averaged background quantities in the $\zeta$ action.  In \cite{ali1}, only the loops arising from the interaction potential have been considered. Here, we determine all cubic interactions involving $\chi$ and $\zeta$, and estimate the total one loop correction to the $\zeta$ power spectrum. Not surprisingly, our computations confirm the findings of \cite{ali1} and show a significant contribution to the  $\lf\zeta\zeta\rg$  correlation function. 

Our second aim in this work is to determine the  $\chi$ loop contributions to the non-gaussianity and to the tensor power spectrum (in single field models, the bi-spectrum is not altered by the parametric resonance effects \cite{yeni1}). Notable modifications to the scalar power spectrum found in \cite{ali1} indicate the existence of similar significant corrections for these observables. We estimate the amount of non-gaussianity from the $\zeta$-three point function by calculating the one loop graphs arising from the cubic interactions. It turns out that these corrections to the three point function can be expressed in terms of the two point function and it is possible to read the shape independent non-gaussianity order parameter $f_{NL}$. Similarly, the $\chi$ field  coupled to the tensor fluctuations yield loop corrections to the tensor power spectrum. Since the tensor field behaves like a test field propagating on the background, the tensor calculation is not affected by different time slices of spacetime. We find that the tensor power spectrum is moderately corrected by the loops in reheating. 

The organization of the paper is as follows. In the next section, we consider the chaotic $m^2\f^2$ model with the extra preheating scalar field $\chi$ to which the inflaton decays in the parametric resonance regime. 
In section \ref{iii},  we  determine the cubic interaction terms involving the curvature perturbation $\zeta$, the tensor mode $\cc_{ij}$  and the preheating scalar $\chi$.  We then calculate one loop corrections to the scalar power spectrum $\lf\zeta\zeta\rg$, the three-point function $\lf\zeta\zeta\zeta\rg$, the tensor power spectrum $\lf\cc\cc\rg$ and make order of magnitude estimates of these corrections by using the theory of preheating. In section \ref{sec3}, we further consider some higher order interactions involving two $\chi$ fields and determine their loop effects. In \ref{sec4} we conclude with  remarks and future directions. 

\section{The model and linearized fluctuations}
 
\subsection{The background}

Let us consider the chaotic model that has the following potential
\be\label{pot} 
V(\phi,\chi)=\fr12 m^2\phi^2+\fr12 g^2 \phi^2\chi^2,
\ee
where $\f$ is the inflaton and $\chi$ is the reheating scalar, which are propagating in a flat FRW background 
\be
ds^2=-dt^2+a^2(dx^2+dy^2+dz^2).
\ee
This model can be seen to be the prototype of the chaotic inflationary paradigm and preheating. As it is well known, a period of inflation can be realized if initially $\f>\tilde{M}_p$ and the nearly exponential expansion ends roughly when  $\phi\sim \tilde{M}_p/20$ (see e.g. \cite{reh4}), where $\tilde{M}_p^2=1/G$ (we define $M_p$ to be the reduced Planck mass $M_p^2=1/8\pi G$). 

During inflation and in the first stage of preheating where the backreaction effects are negligible, the $\chi$ background vanishes
\be
\chi=0. 
\ee
Following the exponential expansion, $\f$ starts oscillating about its minimum $\f=0$. Assuming 
\be
m\gg H,
\ee
which is generically satisfied in this model, the background field equations can be approximately solved as
\be\label{b1}
\phi(t)\simeq\Phi \sin(mt),\hs{5}a=a(t),
\ee
where 
\be \label{fh}
a\simeq\left(\fr{t}{t_R}\right)^{2/3},\hs{5}\Phi\simeq\fr{\F_0}{mt},\hs{5}H\simeq\fr{2}{3t}.
\ee
Note that the amplitude obeys $\dot{\Phi}+3H\Phi/2\simeq0$, where the dot denotes the time derivative. 

We define $t_R$ and $t_F$ as follows: 
\bea
&&t_R:\textrm{Beginning of reheating,}\nn\\
&&t_F:\textrm{End of the first stage of preheating.}\nn
\eea
After  the time $t_F$, the $\chi$ particles created out of the vacuum start affecting the background and thus the backreaction effects are set in. Our aim is to calculate the $\chi$ loop corrections to the cosmological correlation functions, which are effective in the time interval $(t_R,t_F)$. 

Some features of preheating depend on the parameters of the model and many cases are discussed numerically in \cite{reh4}. For our estimates, we will use the following canonical set that gives the broad parametric resonance:
\be\label{s1}
m=10^{-6}\tm\hs{10}g=10^{-2}.
\ee 
In that case, the first stage of preheating ends after about $11$ inflaton oscillations and one has \cite{reh4}
\be\label{s2}
H(t_F)\simeq 10^{-2} m \hs{5}\Phi(t_F)=5\times 10^{-3}\tm.
\ee
One may also note that
\be\label{tn}
mt_R\simeq 1,\hs{7}mt_F\simeq \fr{200}{3},
\ee
which can be determined from $m/H(t_F)=3mt_F/2\simeq 100$ and the fact that the first stage ends after $11$  oscillations. The initial amplitude in \eq{fh} is given by $\F_0\simeq \tm/20$. 

In the physical momentum space, the first resonance band is given by $q_{phys}\in(0,q_*)$ where 
\be\label{pi}
q_*=\sqrt{gm\Phi}.
\ee
In general, there are other resonance bands which can be important for preheating  \cite{reh4}. In the model we are studying, the first instability band gives the largest contribution and in the following we simply underestimate the loop corrections by neglecting the effects  of other resonance bands. The $\chi$ momentum modes sitting in the band $(0,q_*)$ encounter exponential amplification. In determining $q_*$, we will use the smallest value of $\Phi$, i.e $\Phi(t_F)$ in \eq{s2}. The first stage ends when the interaction potential energy density $g^2\f^2\chi^2$ becomes comparable to the inflaton potential energy density $m^2\f^2$, since after that moment the frequency of the inflaton oscillations are affected by the $\chi$ particles. This implies \cite{reh4} 
\be\label{ce}
\lf \chi(t_F)^2\rg  \simeq \fr{m^2}{g^2}.
\ee
As we will see below, \eq{ce} is important for estimating  the $\chi$ loop corrections. 

In a generic two-field model, the adiabatic field $\s$ and the entropy perturbation $\d s$ are defined by \cite{ent} 
\bea
&&\dot{\sigma}=(\cos\th)\df+(\sin\th)\dot{\chi},\\
&&\d s=(\cos\th)\d\chi-(\sin\th)\d \f,
\eea
where
\be
\cos\th=\fr{\df}{\sqrt{\df^2+\dot{\chi}^2}},\hs{3} \sin\th=\fr{\dot{\chi}}{\sqrt{\df^2+\dot{\chi}^2}}. 
\ee
Since the background value of $\chi$ is zero,  we have $\th=0$, $\s=\f$ and $\d s=\d \chi$, which shows that in this model $\f$ is the adiabatic mode and $\chi$ is the entropy mode. 

\subsection{Quadratic actions and  mode functions}

The full action governing the dynamics of the system can be written in  the ADM form as (we set $M_p=1$) 
\be\label{a}
S=\fr12 \int \sqrt{h} \left[NA+\fr{B}{N}\right],
\ee
where $N$ and $N^i$ are the standard lapse and shift functions  of the metric
\be
ds^2=-N^2dt^2+h_{ij}(dx^i+N^i dt)(dx^j + N^j dt),
\ee
$K_{ij}=\fr12(\dot{h}_{ij}-D_i N_j-D_j N_i)$, $K=h^{ij}K_{ij}$, $D_i$ is the derivative operator of $h_{ij}$ and 
\bea
&&A=R^{(3)}-2V-h^{ij}\del_i\phi\del_j\f-h^{ij}\del_i\chi\del_j\chi,\label{aa}\\
&&B=K_{ij}K^{ij}-K^2+(\dot{\f}-N^i\del_i\f)^2+(\dot{\chi}-N^i\del_i\chi)^2.\label{bb}
\eea
We define the perturbations as 
\bea
&&h_{ij}=a^2e^{2\zeta}(e^\cc)_{ij},\\
&&\chi=0+\chi,\label{cpert}
\eea
where the gauge is completely fixed by imposing
\be\label{gauge}
\vf=0,\hs{5}\del_i\cc_{ij}=0,\hs{5} \cc_{ii}=0.
\ee
Here, $\vf$ denotes the inflaton fluctuation and in this gauge the inflaton takes its background value $\phi=\phi(t)$ given in \eq{b1}.  Note that we use the same letter $\chi$ to denote the reheating scalar fluctuation in \eq{cpert} since the background value of $\chi$ vanishes. As pointed out in \cite{ekw}, the lapse $N$ can be solved exactly as $N^2=B/A$. However, to determine the action up to cubic order it is enough to solve the constraints to linear order, which gives \cite{mal} 
\be\label{ls}
N=1+\fr{\dot{\zeta}}{H},\hs{5}N^i=\d^{ij}\del_j\psi,\hs{5}\psi=-\fr{\z}{a^2H}+\fr{\dot{\f}^2}{2H^2}\del^{-2}\dot{\z}.
\ee 
Note that neither $\chi$ nor $\cc_{ij}$ appear in the solutions of $N$ and $N^i$ to this order.  By expanding the action \eq{a}, one may  obtain the following well known quadratic actions 
\bea
&&S_\z^{(2)}=\fr12\int a^3\fr{\dot{\f}^2}{H^2}\left[\dot{\z}^2-\fr{1}{a^2}(\del\zeta)^2\right],\nn\\
&&S_\chi^{(2)}=\fr12\int a^3\left[\dot{\chi}^2-\fr{1}{a^2}(\del\chi)^2-g^2\f^2\chi^2\right],\label{qac}\\
&&S_\cc^{(2)}=\fr18\int a^3\left[\dot{\cc}_{ij}^2-\fr{1}{a^2}(\del\cc_{ij})^2\right],\nn
\eea
which are valid both during inflation and reheating. The $\z$ kinetic term vanishes at times when $\dot{\f}=0$ and the $\z$ propagator diverges at these times. This divergence must be cured to make the loop contributions well defined. 

The free fields can be expanded as 
\bea
&&\z=\fr{1}{(2\pi)^{3/2}}\int d^3k\, e^{i\vec{k}.\vec{x}}\,\z_k(t) a_{\vec{k}}+h.c.\label{m1}\\
&&\chi=\fr{1}{(2\pi)^{3/2}}\int d^3k\, e^{i\vec{k}.\vec{x}}\,\chi_k(t) \tilde{a}_{\vec{k}}+h.c.\label{m2}\\
&&\cc_{ij}=\fr{1}{(2\pi)^{3/2}}\int d^3k\, e^{i\vec{k}.\vec{x}}\,\cc_k(t) \e_{ij}^s \tilde{a}^s_{\vec{k}}+h.c.\nn
\eea
where $s=1,2$ and the ladder operators obey the usual commutator relations, e.g. $[a_k,a^\dagger_{k'}]=\d^3(k-k')$. The polarization tensor $\e^s_{ij}$ has the following properties
\be
k^i\e^s_{ij}=0,\hs{5} e^s_{ii}=0,\hs{5} \e^s_{ij}e^{s'}_{ij}=2\d^{ss'}. 
\ee
To satisfy the canonical commutation relations, the mode functions must obey the Wronskian conditions 
\bea
&&\z_k\dot{\z}_k^*-\z_k^*\dot{\z}_k=\fr{H^2i}{a^3\dot{\f}^2},\nn\\
&&\chi_k\dot{\chi}_k^*-\chi_k^*\dot{\chi}_k=\fr{i}{a^3},\label{w}\\
&&\cc_k\dot{\cc}_k^*-\cc_k^*\dot{\cc}_k=\fr{4i}{a^3}.\nn
\eea
On the other hand, the linearized mode equations become  
\bea
&&\ddot{\z}_k+\left[3H+2\fr{\ddot{\f}}{\dot{\f}}-2\fr{\dot{H}}{H}\right]\dot{\z}_k+\fr{k^2}{a^2}\z_k=0,\nn\\
&&\ddot{\chi}_k+3H\dot{\chi}_k+\left[g^2\f^2+\fr{k^2}{a^2}\right]\chi_k=0,\label{le}\\
&&\ddot{\cc}_k+3H\dot{\cc}_k+\fr{k^2}{a^2}\cc_k=0.\nn
\eea
Note that the equation for $\chi_k$ gets a contribution from the potential \eq{pot}, which is responsible for the parametric resonance. 

We will be interested in the superhorizon $\z_k$ and $\cc_k$ modes. Neglecting the $k^2/a^2$ terms in \eq{le} one can easily obtain two linearly independent {\it superhorizon solutions} which can be written as
\be\label{sh}
\z_k\simeq \left[\zo+c_k f(t)\right],  \hs{5} \cc_k\simeq \left[\cco+d_k g(t)\right],
\ee
where $\zo$, $\cco$, $c_k$ and $d_k$ are constants and 
\be\label{fg}
\fr{df}{dt}=\fr{H^2}{a^3\df^2}, \hs{5}  \fr{dg}{dt}=\fr{1}{a^3}.
\ee
As usual, the modes \eq{sh} have the constant and the decaying pieces, and the normalization conditions in \eq{w} imply 
\be\label{n}
\zo c_k^* -\zo{}^* c_k=i,\hs{5} \cco d_k^* -\cco{}^* d_k=4i.
\ee
One may note the mass dimensions\footnote{Note that $\cc_{ij}$ commutation relation has a factor of $1/M_p^2$ in the right hand side, which is set to one. This is why the mass dimensions of $c_k^0$ and $d_k^0$ are different.} of the constants as $[\zo]=M^{-3/2}$, $[c_k]=M^{3/2}$, $[\cco]=M^{-3/2}$ and $[d_k]=M^{-1/2}$.

To be able to calculate the $\chi$ loop effects, we need to determine the behavior of the $\chi$ modes, especially the ones in the resonance band, in  detail. For that, one may write the mode function in the WKB form as follows
\be\label{est}
\chi_q=\fr{1}{\sqrt{2a^3\o_q}}\left[\a_qe^{-i\int \o_q}+ \b_qe^{i\int \o_q}\right],
\ee
where 
\be\label{wq}
\o_q^2=g^2\f^2+\fr{q^2}{a^2}-\fr94 H^2-\fr32 \dot{H}.
\ee
The Wronskian condition is satisfied by imposing $|\a_q|^2-|\b_q|^2=1$. During inflation, $\chi$ becomes a very massive field with mass $g\F_0$. As a result, for the modes of interest the Bunch-Davies mode function in the beginning of reheating can be written up to an irrelevant phase as
\be\label{ic}
\chi_q(t_R)\simeq \fr{1}{\sqrt{2a^3g\F_0}}. 
\ee
This shows that at the end of the exponential expansion these individual $\chi_q$ modes are suppressed by $a^{-3/2}$  and this is the main reason for the metric preheating scenario of \cite{ca,mph1,mph2} to break down, as it is discussed in \cite{mpk1,mpk2,mpk3} (it is possible to circumvent this suppression in some models, as it is shown in \cite{mph4,mph5,mph6}). During preheating, $\chi_q$ changes non-adiabatically as the inflaton passes through the potential minimum $\f=0$. This process can be formulated as the particle creation by parabolic potentials which gives the exponential increase $\b_q = e^{\m_q m t}$ for the modes in the instability bands, where $\m_q$ is an index characterizing the exponential growth. 

From \eq{est} one may find that 
\be
|\chi_q|^2=\fr{1}{2a^3\o_q}\left[1+2|\b_q|^2+2\textrm{Re}\left(\a_q\b_q^*e^{-2i\int \o_q}\right)\right].
\ee
For $|\a_q|\simeq|\b_q|\gg1$, it is possible to see that $|\chi_q|^2$ oscillates between $1/(2a^3\o_q)$ and $4|\b_q|^2/(2a^3\o_q)$ with the frequency $\o_q$. To determine the phase of $\chi_q$, one may define $\th_q$ as
\be\label{c-amp}
\chi_q=|\chi_q|e^{-i\th_q}.
\ee
Then, the Wronskian condition \eq{w} gives
\be\label{pep}
\fr{d\th_q}{dt}=\fr{1}{2a^3|\chi_q|^2}, 
\ee
i.e. up to an unimportant constant the phase is uniquely fixed by the amplitude $|\chi_q|$. 

The growth of the modes in the first instability band can be described by introducing an effective index $\m_q\simeq \m$, and for the parameters given in \eq{s1} one has \cite{reh4}
\be\label{apin}
\m\simeq 0.13.
\ee
Since $|\chi_q|^2\propto |\b_q|^2$, the amplitude $|\chi_q|$ can be seen to be enlarged by a factor of $\exp(0.13\times 2 \pi)=2.26$, after each oscillation.  

To estimate the magnitude of the amplitude $|\chi_q|$ at the end of the first stage of preheating, one may look at the expectation value $\lf \chi^2\rg$, which is given by 
\be\label{34} 
\lf \chi^2\rg=\fr{1}{(2\pi)^3}\int d^3 q |\chi_q|^2 \simeq \fr{4\pi}{(2\pi)^3}a^3 q_*^3 |\chi_{q_*}|^2
\ee
where in the last equality we restrict the momentum integral to the first (and the most important) instability band, which is supposed to give the dominant contribution to the vacuum expectation value; we switch to the physical momentum space and introduce  $|\chi_{q_*}|$ to denote a mean value for the modes in this instability band. Note that the $4\pi$ factor in \eq{34} comes from the angular directions in the momentum space. Comparing with \eq{ce} one may deduce that at the end of the first stage  
\be\label{cestonce}
|\chi_{q_*}|_{max}^2\simeq 2\pi^2\fr{m^2}{a^3q_*^3g^2}.
\ee
As pointed out above, the amplitude $|\chi_q|$ is actually an oscillating function that has frequency $\o_q$. However, one has $\o_{q_*}\gg m$  and thus $|\chi_{q_*}|$ oscillates much faster than the background inflaton field. As a result,  \eq{cestonce} should be divided by 2 to give a time averaged value for the amplitude. We also use the index $\mu$ to obtain the amplitude in the middle of the period and define
\be\label{cest}
|\chi_{q_*}|^2\simeq \fr{\pi^2m^2}{a^3q_*^3g^2}\,e^{-2\pi\m}.
\ee
The phase corresponding to \eq{cest}  can determined from \eq{pep} as 
\be\label{cestp} 
\th_{q_*}\simeq \fr{q_*^3g^2}{2\pi^2m^2}\,e^{2\pi\mu}\,t. 
\ee
These estimates will be crucial in determining the strength of a graph in the in-in perturbation theory. 

We define the scalar and the tensor power spectra in the momentum space, i.e. $P_k^\z$ and $P_k^\cc$, from the two point functions in the form 
\bea\label{sp}
&&\lf\z(t,\vec{x})\z(t,\vec{y})\rg=\fr{1}{(2\pi)^{3}}\int d^3k \,e^{i\vec{k}.(\vec{x}-\vec{y})}\,P_k^\z(t),\\
\label{tp}
&&\lf\cc_{ij}(t,\vec{x})\cc_{kl}(t,\vec{y})\rg=\fr{1}{(2\pi)^{3}}\int d^3k \,e^{i\vec{k}.(\vec{x}-\vec{y})}\,P_k^\cc(t)\Pi_{ijkl},
\eea
where the polarization tensor $\Pi_{ijkl}$, which is defined as
\be\label{pt}
\Pi_{ijkl}=e^s_{ij}e^s_{kl},
\ee  
obeys $\Pi_{ijkl}\Pi_{klmn}=2\Pi_{ijmn}$. The tree level standard results can be read from \eq{sh} as
\be\label{3l}
P_k^{\z(0)}(t)=|\zo|^2,\hs{5} P_k^{\cc(0)}(t)=|\cco|^2.
\ee
The constants $\zo$ and $\cco$ can be determined from the mode functions of the free fields during inflation and as it is well known they depend on the horizon crossing time for a given $k$ (see \cite{kemal} for a study of loop corrections to the mode functions during inflation).  

\subsection{Smoothing out  spikes of $\zeta$}

In finding $f(t)$ from \eq{fg}, an infinity arises when  the limits of the integration contains a moment giving $\dot{\f}=0$. To avoid these singularities one may try to fix $f(t)$ by an indefinite integral since one only needs a function whose derivative gives \eq{fg}. However, the function obtained in this way is unavoidably singular at times when $\df=0$. Moreover, the loop corrections turn out to involve the time integrals of $f(t)$ or $df/dt$, and these also diverge when $f(t)$ obeys \eq{fg}. 

This pathologic behavior arises due to the bad choice of gauge.\footnote{To avoid this problem, one can use the inflaton fluctuation $\vf$ as the main dynamical variable to calculate the loop quantum corrections and gauge transform to $\z$ at the end of the reheating stage.  See the appendix of \cite{ali1} for an example of how gauge transformations change the time integrals in the loops.}  Namely, $\vf=0$ gauge breaks down at times when $\df=0$ giving rise to the spikes of $\z$. This has already been noted in some earlier work, see e.g. \cite{coh2,brz}.  As discussed in \cite{brz}, although $\z$ becomes an ill defined variable in reheating, $(1+w)\z$ becomes well defined, where $w$ is the equation of state parameter. In our model $1+w=2\df^2/(\df^2+m^2\f^2)$.  

To smooth out the spikes of $\z$, we first note that the Einstein's equations for the background give 
\be\label{fav0}
\df^2=-2M_p^2\dot{H},
\ee
where we display the Planck mass dependence for later use. Since we use \eq{fh} to approximate the Hubble parameter $H$, one may define $\df^2_{\textrm{av}}$ by using \eq{fh} in \eq{fav0} that yields 
\be\label{fav}
\df^2_{\textrm{av}} =\fr{4M_p^2}{3t^2}.
\ee
It is clear that $\df^2_{\textrm{av}}$ gives the ``time" average of the oscillating function $\df^2$. To make $\z$ well defined, one may now replace $\df^2$ by $\df^2_{\textrm{av}}$ in the free action of $\z$ in \eq{qac}. In the  context of the discussion carried out in \cite{brz}, this is equivalent to using an average equation of state parameter $w_{\textrm{av}}$ instead of the actual one. Consequently, one simply treats the $\z$ variable as if it evolves in a matter dominated universe. In that case, the new function obeys
\be\label{fgn}
\fr{df}{dt}=\fr{H^2}{a^3\df^2_{\textrm{av}}}=\fr{1}{3M_p^2a^3}.
\ee
A simple integration then gives  
\be\label{fs}
f\simeq\fr{2}{9M_p^2Ha^3}, \hs{5}g\simeq \fr{2}{3Ha^3},
\ee
where we use \eq{fgn} and \eq{fg} for $f(t)$ and $g(t)$, respectively. As we will see below, the loop contributions turn out to depend on the difference of two $f(t)$ or the difference of two $g(t)$ functions, and therefore there is no need to fix the  integration constants in \eq{fs}. 

\section{Cubic interactions and loop corrections} \label{iii} 

Using \eq{cpert} and \eq{ls} in \eq{a}, a straightforward calculation gives the following cubic action involving two $\chi$ fields:
\bea
S^{(3)}=&&\fr12\int a^3\left[-3g^2\f^2\z\chi^2-\fr{g^2\f^2}{H}\dot{\z}\chi^2-\fr{1}{a^2}\z(\del\chi)^2-\fr{1}{a^2H}\dot{\z}(\del\chi)^2-\fr{1}{H}\dot{\z}\dot{\chi}^2+3\z\dot{\chi}^2-2N^i\dot{\chi}\del_i\chi\right. \nn \\
&&\left. +\fr{1}{a^2}\cc^{ij}\del_i\chi\del_j\chi\right]. \label{a3} 
\eea
Combining this cubic action with the quadratic ones given in \eq{qac} and switching to the Hamiltonian formulation,  one may find the cubic interaction Hamiltonian containing two $\chi$ fields as 
\be\label{h3}
H_I^{(3)}=\int d^3 x\,a^3 \left[\dot{\z} O_1+\z O_2+\cc^{ij}O_{ij}\right],
\ee
where
\bea
&&O_1=\fr{g^2\f^2}{2H}\chi^2+\fr{1}{2Ha^2}(\del\chi)^2+\fr{1}{2H}\dot{\chi}^2-\fr{\df^2}{2H^2}\del^{-2}\del_i(\dot{\chi}\del_i\chi),\label{o1}\\
&&O_2=\fr{3}{2}g^2\f^2\chi^2+\fr{1}{2a^2}(\del\chi)^2+\fr{3}{2}\dot{\chi}^2+\fr{1}{a^2H}\del_i(\dot{\chi}\del_i\chi),\label{o2}\\
&&O_{ij}=-\fr{1}{a^2}\del_i\chi\del_j\chi.
\eea
Although it is not indicated explicitly, all the fields appearing in \eq{h3} can be taken to be the interaction picture fields that enter in the in-in perturbation theory as it is formulated  in \cite{w2}. In obtaining \eq{h3} we only spatially integrate by parts the last term in the first line of \eq{a3} to replace the shift $N^i$ by its potential $\psi$ given in \eq{ls}.  

For any given operator $O$, the in-in formalism can be applied to obtain the following perturbative expansion for the vacuum expectation value \cite{w2} 
\be\label{inp}
\lf O(t)\rg=\sum_{N=0}^{\infty} i^N \int_{t_R}^t dt_N\int_{t_R}^{t_N}dt_{N-1}...\int_{t_R}^{t_2}dt_1\left< [H_I(t_1),[H_I(t_2),...[H_I(t_N),O(t)]...]\right>.
\ee
where the lower limit of the time integrals is set to $t_R$ rather than $-\infty$ since we are interested in the loop effects during reheating. In general, the two terms in a given commutator in \eq{inp} have different $i\e$ prescriptions, which would be important for the convergence of the time integrals if they were extended to $-\infty$. In \eq{inp}, this technical problem does not arise  since the time integrals span a finite time interval. Because the Hamiltonian contains the products of the fields and their time derivatives (i.e. their momenta) there is an ordering ambiguity in \eq{inp}. Although it is crucial to solve this ambiguity to obtain  exact results (for instance by utilizing a symmetric ordering prescription), this will not be a problem for our order of magnitude estimates. 

\subsection{The scalar power spectrum}

\begin{figure}
\centerline{
\includegraphics[width=6cm]{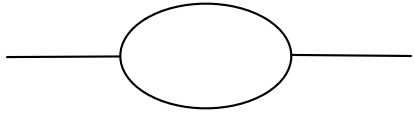}}
\caption{The 1-loop graph arising from the interaction Hamiltonian \eq{h3} that contributes to the $\lf\z\z\rg$ correlation function during reheating. The graph schematically indicates the vertices coming from the interaction Hamiltonian and possible contractions or commutators of the external $\z$ fields and the internal $\chi$ fields giving rise to the loop. One may draw similar graphs with time or spatial derivatives acting on the fields. The disconnected graphs, where the two $\chi$ fields in the same interaction vertex are contracted with each other, are suppressed.}
\label{fig1}
\end{figure}

We first calculate the one loop correction to the scalar power spectrum arising from the cubic interaction Hamiltonian \eq{h3}. Since $H_I^{(3)}$ is linear in $\z$, the first nonzero contribution in \eq{inp} appears for $N=2$ and the corresponding terms can be pictured like the graph in Fig. \ref{fig1}. Since  $H_I^{(3)}$ contains two $\chi$ fields and a volume factor of $a^3$, the suppression of the $\chi_q$ mode by $a^{-3/2}$ is compensated in the interaction Hamiltonian. On the other hand, the three dimensional  loop integral  must be converted to the physical momentum space since the instability band is given in the physical scale  in \eq{pi}. This yields an extra enlargement factor of $a^3$.

Using \eq{inp} for the operator $\z(t,\vec{x})\z(t,\vec{y})$ with $N=2$ gives the following vacuum expectation values of the nested commutators:
\bea
&&\left<\left[\z(t_1,\vec{z}_1)O_2(t_1,\vec{z}_1),\left[\z(t_2,\vec{z}_2)O_2(t_2,\vec{z}_2),\z(t,\vec{x})\z(t,\vec{y})\right]\right] \right>, \label{most0}\\
&&\left<\left[\z(t_1,\vec{z}_1)O_2(t_1,\vec{z}_1),\left[\dot{\z}(t_2,\vec{z}_2)O_1(t_2,\vec{z}_2),\z(t,\vec{x})\z(t,\vec{y})\right]\right]\right>,\label{most}\\
&&\left<\left[\dot{\z}(t_1,\vec{z}_1)O_1(t_1,\vec{z}_1),\left[\z(t_2,\vec{z}_2)O_2(t_2,\vec{z}_2),\z(t,\vec{x})\z(t,\vec{y})\right]\right]\right>,\label{most1}\\
&&\left<\left[\dot{\z}(t_1,\vec{z}_1)O_1(t_1,\vec{z}_1),\left[\dot{\z}(t_2,\vec{z}_2)O_1(t_2,\vec{z}_2),\z(t,\vec{x})\z(t,\vec{y})\right]\right]\right>.\label{most2}
\eea
From the identity $[AB,C]=A[B,C]+[A,C]B$, one sees that there are terms either containing $\lf \z\z\rg [\z,\z]$ or  $[\z,\z] [\z,\z]$ (or similar terms where two of the $\z$'s are replaced by $\dot{\z}$). Each commutator $[\z,\z]$ or $[\dot{\z},\z]$  yields a factor of  $a^{-3}$. As pointed out above, there is  one $a^3$ factor coming from the loop momentum integral, which may compensate a single $a^{-3}$. This shows that the terms involving two  commutators $[\z,\z] [\z,\z]$ are suppressed. Similarly, the expectation value $\lf\z\dot{\z}\rg$  also gives an extra factor of $a^{-3}$ since the time derivative  kills the constant piece in \eq{sh}, therefore these  are also suppressed. 

From \eq{fs} one observes that $df/dt\simeq Hf$. Besides,  while the commutator $[\z,\z]$ gives the function $f(t)$ the commutator $[\z,\dot{\z}]$ yields the function $df/dt$. Namely, from the mode expansion \eq{m1} one easily  calculates
\bea
&&\left[\z(t_2,\vec{z}_2),\z(t,\vec{x})\right]=\fr{1}{(2\pi)^3}\int d^3 k e^{i\vec{k}.(\vec{z}_2-\vec{x})}\left[\z_k(t_2)\z_k^*(t)-\z_k^*(t_2)\z_k(t)\right],\\
&&\left[\dot{\z}(t_2,\vec{z}_2),\z(t,\vec{x})\right]=\fr{1}{(2\pi)^3}\int d^3 k e^{i\vec{k}.(\vec{z}_2-\vec{x})}\left[\z_k(t_2)\dot{\z}_k^*(t)-\z_k^*(t_2)\dot{\z}_k(t)\right]. 
\eea
Using \eq{sh} in these  commutators, we see that for superhorizon modes  the first commutator gives $[f(t_2)-f(t)]$ and the second one yields $df(t_2)/dt$ in the square brackets. From these observations and using  $O_1$ and $O_2$ given in \eq{o1} and \eq{o2}, one may conclude that all the terms in \eq{most0}-\eq{most2} have the same order of magnitude. However, since $f(t)$ is a slowly varying function and moreover $[f(t_2)-f(t)]$ vanishes when $t_2=t$, we find that  the loop corrections containing the commutator $[\z,\dot{\z}]$ is larger than the corrections with the commutator $[\z,\z]$. To sum up, we find that the largest of all the terms that arise in \eq{inp} is the one coming from \eq{most} that has the structure $\lf \z\z\rg [\z,\dot{\z}]$. Defining the function $F(t_1,t_2,k)$ by 
\be\label{oo} 
\left<O_2(t_1,\vec{z}_1)O_1(t_2,\vec{z}_2)\right>=\fr{1}{(2\pi)^3}\int d^3 k\, e^{i\vec{k}.(\vec{z}_1-\vec{z}_2)}\,F(t_1,t_2,k),
\ee
and using \eq{sh}, \eq{n} and \eq{sp}, one can determine the largest correction as 
\be\label{cor1}
P_k^\z(t_F)^{(1)}\simeq P_k^{\z(0)}\,(2i)\int_{t_R}^{t_F}dt_2\int_{t_R}^{t_2}dt_1\, a(t_1)^3\,a(t_2)^3\,\fr{df}{dt}(t_2)\left[F(t_1,t_2,k)-c.c.\right],
\ee
where $k$ denotes  the comoving cosmological superhorizon scale of interest and $t_F$ marks the end of the first stage of preheating as defined above. It is remarkable that the one-loop correction $P_k^\z(t_F)^{(1)}$ becomes  a multiple of the the tree level function $P_k^{\z(0)}$ given in \eq{3l}. From \eq{o1} and \eq{o2}, $F(t_1,t_2,k)$ can be found as 
\be\label{49} 
F(t_1,t_2,k)=\fr{3g^4}{2(2\pi)^3H(t_2)}\f^2(t_1)\f^2(t_2)\int d^3 q\, \chi_{q}(t_1)\chi_{k+q}(t_1)\chi_{q}^*(t_2)\chi_{k+q}^*(t_2)+...
\ee
where only the contribution of the first terms in \eq{o1} and \eq{o2} are written explicitly. 

The momentum integral in \eq{49}, and similar loop integrals below, do diverge and these must be regularized/renormalized before making any order of magnitude estimates. To figure out the contribution of the modes in the resonance band and for regularization, we simply cutoff the integral in \eq{49} with $aq_*$,  where $q_*$ is given by \eq{pi}. It is easy to see that this procedure corresponds to the adiabatic regularization where one uses the WKB mode function \eq{est} and discard the pieces with $\a_q$ that give infinities.\footnote{The same regularization has been used in \cite{reh4} to determine  the parametric resonance effects. Therefore,  using the WKB regularization for our loop corrections is crucial for consistency since we heavily use the results of \cite{reh4} in our estimations.} Note that since $|\a_k|\to 1$ and $|\b_k|\to 0$ as  $k\to\infty$, adiabatic regularization guarantees the finiteness of the loop integrals. Initially we have $\a_q(t_R)=1$ and $\b_q(t_R)=0$; and $\b_q$ increases with time in the resonance band and stays vanishingly small for high energy modes since they propagate adiabatically. Therefore, using the resonance scale for the momentum cutoff is equivalent to the adiabatic regularization. 

On the other hand, from  \eq{s1} and \eq{s2} one sees that $q_*\simeq 10^{-6}\tm\ll \tm$.  Consequently, in a standard renormalization procedure that is more systematic than the simple adiabatic regularization, the UV subtractions should not change our estimates since the cutoff scale $q_*$ corresponds to a relatively low energy scale. Indeed, it is not difficult to convince oneself that the adiabatic subtractions that is automatically performed by our momentum cutoff must be the same with the UV subtractions, i.e. the result obtained with our cutoff must be the same with the finite result obtained after UV subtractions. To see this, imagine that the loop integral is regularized by a UV cutoff $\Lambda\sim M_p$. Then, our method is equivalent to throwing out the momentum range $(q_*,\Lambda)$, which can be thought to be canceled out by the $\Lambda$-dependent counterterms. In this procedure, the finite renormalizations can be fixed by referring to the tree level inflationary results. Note that the dimensional regularization is very difficult to implement in this computation since the exact form of the mode function $\chi_q$ is not known. 

The correction \eq{cor1} modifies not only the amplitude but also the index of the power spectrum. This nontrivial $k$-dependence ensures that \eq{cor1} cannot be interpreted as a finite renormalization effect. On the other hand,  the change in the index turns out to be small for cosmologically interesting scales\footnote{The index is meaningfully modified for the modes entering the horizon during reheating that may change the primordial black hole formation, see  \cite{bh}.}  since in that case $k\ll a q_*$. Therefore, the $k$ dependence of \eq{cor1} is negligible and to a very good approximation one may ignore it by setting $k=0$. 

As discussed above, $a(t_1)^3$ and $a(t_2)^3$ terms cancel out the scale factor suppressions of the four $\chi_q$ modes. The $1/a^3$ factor that appears in $df/dt$ in  \eq{fgn} can be used to convert the comoving momentum integral in \eq{49} to the physical scale. Thus, all the scale factors in \eq{cor1} simply cancel out each other. 

In what follows we estimate \eq{cor1} to determine the size of the loop effects in reheating. We first focus on the term that is explicitly shown in \eq{49} and then confirm that others give similar contributions. Since the resonant $\chi$ modes encounter most of their growth near the end of the first stage, one may focus on the last  inflaton oscillation for the time integrals in \eq{cor1}, namely, the lower and the upper limits can be set to $mt_F-2\pi$ and $mt_F$, respectively. Using \eq{c-amp},  the square brackets in \eq{cor1} yields the following factor 
\be\label{f5}
\sin\left[\th_q(t_1)+\th_{k+q}(t_1)-\th_q(t_2)-\th_{k+q}(t_2)\right].
\ee
We see that the leading order contribution does not cancel out since the phase factors have different time arguments. In \eq{49}, there are four $\chi_q$ modes integrated out in the first instability band, which can be estimated as $q_*^3|\chi_{q_*}|^4$, where  $|\chi_{q_*}|$ is the mean value of the modes introduced in \eq{cest}. The function $df/dt$ can be read from \eq{fgn}.  Treating  the slowly changing factors like $H$ and $\Phi$ as constants one finally  finds that 
\be\label{e1} 
P_k^\z(t_F)^{(1)}\simeq  P_k^{\z(0)}\left[\fr{24\pi}{(2\pi)^3}\right]\left[\fr{g^4\Phi(t_F)^4}{H(t_F)}\right]\left[q_*^3\left|\chi_{q_*}\right|^4\right]\left[\fr{C_1}{m^2}\right]\left[\fr{1}{3M_p^2}\right]+...
\ee
where the dimensionless constant $C_1$ is given by
\be \label{time1}
C_1=\int_{mt_F-2\pi}^{mt_F}mdt_2\int_{mt_F-2\pi}^{mt_2}mdt_1\sin^2(mt_1)\sin^2(mt_2)\sin\left[2\th_{q_*}(t_1)-2\th_{q_*}(t_2)\right].
\ee
Recall that the phase $\th_{q_*}$ is defined in \eq{cestp}. 

For our set of parameters \eq{s1}, the constant $C_1$ can be determined by a numerical integration that yields $C_1\simeq 0.078$. Using then \eq{s1}, \eq{s2}, \eq{pi} and \eq{cest} in \eq{e1}, we obtain 
\be\label{e2}
P_k^\z(t_F)^{(1)}\simeq  (1.1+...)\, P_k^{\z(0)},
\ee
which becomes larger than the tree level contribution. The Planck mass suppression of  \eq{e1} is compensated by many different factors. The smallest mass scale in the problem, i.e. the Hubble parameter, shows up in the denominator because of the interaction term  \eq{o1}. The mass of the inflaton $m$, the second smallest, also appears in the denominator. On the other hand, the background inflaton amplitude $\Phi$, which is moderately smaller than $M_p$, appears in the numerator with power four due to the first two terms in the interactions \eq{o1} and \eq{o2}. Finally, the mode function $\chi_q$ is amplified exponentially, which also helps the growth considerably. Therefore, different ingredients of this chaotic model play crucial roles for overcoming the Planck mass suppression. 

Let us now consider the contributions of the other terms in \eq{49}, which can be determined from the definition \eq{oo}. From \eq{o1} and \eq{o2}, these consist of the products of four $\chi$ fields, on which certain time or spatial derivatives act (there is also a nonlocal term with $1/\del^{2}$ that involves the Green function of the Laplacian).  In \eq{cor1}, only the imaginary part of $F(k_1,k_2)$  appears in the square brackets. One can easily see that after taking the imaginary part, each product  yields a term similar to \eq{f5} and thus the leading order contributions do not cancel out. On the other hand, the time integrals are very similar to \eq{time1} and they can all be estimated to give $C/m^2$. One may also note that a partial derivative $\del_i$ would produce $q_i$ in momentum space and $\dot{\chi}_q\simeq \o_q \chi_q$, where $\o_q$ is given in \eq{wq}.  Therefore, to estimate the size of a correction one may simply replace $g^2\Phi^2$ factor in the second square bracket in \eq{e1}, which arises due to $g^2\f^2$  terms in \eq{o1} and \eq{o2}, by $\o_{q_*}^2$ corresponding to $\dot{\chi}^2$ or $q_*^2$ corresponding to $(\del\chi)^2$ (note that $1/a^2$ factor, which multiplies $(\del\chi)^2$ in \eq{o1} and \eq{o2},  converts the comoving momentum scale arising from the spatial partial derivative to the physical momentum scale). Similarly, the magnitudes of the nonlocal terms can be estimated by using the Green function for the Laplacian and the correlation length corresponding to the $\chi$ fluctuations, which is roughly equal to $1/q_*$ as shown in \cite{ali2}. In all these different cases one may see that the contributions have the same order of magnitude with \eq{e2}, since for our numerical choice of parameters \eq{s1} one has $g\Phi\simeq \o_{q_*}\simeq 7q_*$. The sign of each contribution depends on the phases through the expressions like \eq{f5}, which is sensitive to the initial conditions \cite{reh4}. In any case, one deduces from \eq{e2} that 
\be\label{e222}
P_k^\z(t_F)^{(1)}\simeq {\cal O}(10)\, P_k^{\z(0)},
\ee
since there are 16 similar contributions.  Eq. \eq{e222}  is consistent with the estimates given in \cite{ali1}. 

Because the one loop correction \eq{e2} is  larger than the tree level result, the in-in perturbation theory might be broken down in this model. Since the modes of the $\chi$ field is exponentially amplified during preheating, the quantum corrections are enlarged when more $\chi$ fields circulate in the loops. As we will see, the results of the next section will support this expectation, i.e. the lower order loop corrections that are supposed to give larger contributions than  \eq{e2} become smaller due to the less number of $\chi$ modes circulating in the loops.  A similar situation also arises for $f_{NL}$ as we will discuss in the next section. 

\subsection{Non-gaussianity}

\begin{figure}
\centerline{
\includegraphics[width=3.5cm]{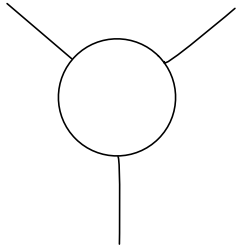}}
\caption{The one loop graph arising from the interaction Hamiltonian \eq{h3} that contributes to the three point function $\lf\z\z\z\rg$. The external and the internal lines correspond to the $\z$ and  the $\chi$ fields, respectively. The time and the spatial partial derivatives acting on the fields are not indicated in the graph.} 
\label{fig2}
\end{figure}

To calculate the non-gaussianity arising from the cubic interaction Hamiltonian \eq{h3}, we express the three point function in the position space as
\be\label{def3}
\left<\z(t,\vec{x})\z(t,\vec{y})\z(t,\vec{z})\right>=\int d^3k_1 d^3k_2 d^3k_3\d(k_1+k_2+k_3)e^{i\vec{k}_1.\vec{x}+\vec{k}_2.\vec{y}+\vec{k}_3.\vec{z}} P(k_1,k_2,k_3).
\ee
The  function $P(k_1,k_2,k_3)$ measures the size of the non-gaussianity involving the comoving superhorizon scales $k_1$, $k_2$ and $k_3$ that obey $k_1+k_2+k_3=0$. To pin down the loop corrections one may use \eq{inp} for $\z(t,\vec{x})\z(t,\vec{y})\z(t,\vec{z})$ and since $H_I^{(3)}$ is linear in $\z$ the first nonzero contribution arises for $N=3$, which gives the diagram in Fig. \ref{fig2}.  

As in the previous subsection, there is one extra enlargement factor of $a^3$ that appears after converting the comoving loop integral to the physical scale. Since the commutator $[\z,\z]$ or $[\dot{\z},\z]$ falls like $1/a^3$, only a single commutator would survive the suppression and all other terms containing two and three $\z$ commutators fall off by the powers of $1/a^3$ and $1/a^6$, respectively (recall that the suppressions of the $\chi$ modes are compensated by $a^3$ factors in the interaction Hamiltonian  $H_I^{(3)}$). Moreover, as it is discussed in detail above, while the $[\z,\z]$ commutator involves the difference of two $f(t)$ functions, the $[\dot{\z},\z]$ commutator yields the function $df/dt$, and the latter gives a larger contribution. Therefore, the biggest one loop correction to $P(k_1,k_2,k_3)$ arises when one uses $\dot{\z}O_1$ in the first and $\z O_2$ in the second and in the third commutators in \eq{inp}.  Repeatedly using the commutator identity $[AB,C]=A[B,C]+[A,C]B$ and defining the function $G(k_1,k_2,k_3)$ as 
\be\label{ngf}
\left<[O_2(t_1,z_1),[O_2(t_2,z_2),O_1(t_3,z_3)]]\right>=\int d^3k_1 d^3k_2 d^3k_3\d(k_1+k_2+k_3)e^{i\vec{k}_1.\vec{x}+\vec{k}_2.\vec{y}+\vec{k}_3.\vec{z}} G(k_1,k_2,k_3),
\ee
one may straightforwardly express the leading order one loop correction in terms of $G(k_1,k_2,k_3) $ as
\be\label{ti} 
P(k_1,k_2,k_3)^{(1)}\simeq-\int_{t_R}^{t_F} dt_3\,a(t_3)^3\int_{t_R}^{t_3}dt_2\,a(t_2)^3\int_{t_R}^{t_2}dt_1\,a(t_1)^3 \fr{df}{dt}(t_3)G(k_1,k_2,k_3)\,P_{k_1}^{\z(0)}P_{k_2}^{\z(0)}+\textrm{cyclic},
\ee
where the extra two terms, which can be obtained by cyclic interchange of momenta, are not written explicitly. 

Using \eq{o1} and \eq{o2} in \eq{ngf}, it is possible to express $G(k_1,k_2,k_3)$ as a loop momentum integral of the mode functions. Indeed a straightforward calculation gives
\bea
&&G(k_1,k_2,k_3)=\fr{9g^6}{2H(t_3)}\f^2(t_1)\f^2(t_2)\f^2(t_3)\fr{1}{(2\pi)^9}\int d^3 q \left[\chi_{q+k_2}(t_2)\chi_{q+k_2}^*(t_3)-c.c\right]\label{g50}\\
&&\left[\left(\chi_{q-k_1}(t_1)\chi_{q-k_1}^*(t_3)-c.c.\right)\left(\chi_q(t_1)\chi_q^*(t_2)+c.c.\right)\right.\nn\\
&&\hs{20}\left.+\left(\chi_{q-k_1}(t_1)\chi_{q-k_1}^*(t_2)-c.c.\right)\left(\chi_q(t_1)\chi_q^*(t_3)+c.c.\right)\right]+...\nn
\eea
where the contributions of the first  terms in  \eq{o1} and \eq{o2}  are expressed explicitly. If $q$ denotes the loop variable that is restricted to the instability band $(0,a q_*)$, again one has $q\gg k_1,k_2,k_3$. Since the modes in the loop integral in \eq{g50} become functions of $q+k_i$, i.e. $\chi_{q+k_i}$, the dependence of $G(k_1,k_2,k_3)$ on its arguments is very weak and one may write  $G(k_1,k_2,k_3)\simeq G$. Using \eq{c-amp} we obtain 
\bea\label{50}
&&|G|\simeq \int d^3q\, \fr{36}{(2\pi)^9}\fr{g^6\Phi^6}{H}|\chi_q|^6\sin^2(mt_1)\sin^2(mt_2)\sin^2(mt_3)\sin\left[\th_q(t_3)-\th_q(t_2)\right]\nn\\
&&\left(\sin[\th_q(t_3)-\th_q(t_1)]\cos[\th_q(t_2)-\th_q(t_1)]+\sin[\th_q(t_3)-\th_q(t_2)]\cos[\th_q(t_3)-\th_q(t_1)]\right)+...
\eea
Since the largest contribution to this loop integral comes when $q$ runs near $aq_*$, one may set $q=aq_*$ and use $\int d^3q\to 4\pi q_*^3$ to estimate the integral. 

It is now possible to use \eq{50} in \eq{ti} to read the three point function. As before, the largest contribution to the time integrals come from the last oscillation period in which $\chi$ modes are amplified most. Keeping the slowly changing factors  like $\Phi$ and $H$ as constants in this last cycle,  we obtain 
\be\label{59}
P^{(1)}(k_1,k_2,k_3)\simeq \fr{144\pi}{(2\pi)^9}\left[q_*^3 \left|\chi_{q_*}\right|^6\right]
\left(\left[\fr{C_2}{m^3}\right]\fr{g^6\Phi(t_F)^6}{H(t_F)}+...\right)\left[\fr{1}{3M_p^2}\right]P_{k_1}^{\z(0)}P_{k_2}^{\z(0)}+\textrm{cyclic},
\ee
where the dimensionless constant $C_2$ is given by 
\bea
&&C_2= \int_{mt_F-2\pi}^{mt_F}m dt_3\,\int_{mt_F-2\pi}^{mt_3}mdt_2\int_{mt_F-2\pi}^{mt_2}mdt_1\sin^2(mt_1)\sin^2(mt_2)\sin^2(mt_3)\sin[\th_{q_*}(t_3)-\th_{q_*}(t_2)]\nn\\
&&\left(\sin[\th_{q_*}(t_3)-\th_{q_*}(t_1)]\cos[\th_{q_*}(t_2)-\th_{q_*}(t_1)]+\sin[\th_{q_*}t_3)-\th_{q_*}(t_2)]\cos[\th_{q_*}(t_3)-\th_{q_*}(t_1)]\right).\label{c2}
\eea
We would like to recall that in this expression the scale factors cancel out each other and the time dependent dimension-full quantities are evaluated at the end of the first stage of preheating. 

The non-gaussianity parameter $f_{NL}$ can be defined as \cite{fnl0,mal} 
\be\label{60}
\z=\z_g-\fr{3}{5}f_{NL} \z_g^2,
\ee
where $\z_g$ denotes the corresponding free quantum field. This definition introduces a shape independent parameter that gives an overall order of magnitude estimate for the scalar non-gaussianity. Calculating the three point function by using \eq{60} and comparing with \eq{59} one finds 
\be \label{fnl0} 
f_{NL}\simeq \fr{240\pi}{(2\pi)^3} \left[q_*^3 \left|\chi_{q_*}\right|^6\right]
\left(\left[\fr{C_2}{m^3}\right]\fr{g^6\Phi(t_F)^6}{H(t_F)}+...\right)\left[\fr{1}{3M_p^2}\right]. 
\ee
For our canonical set \eq{s1}, $C_2$ can be found by a numerical integration that gives $C_2\simeq 0.00057$ (recall that $\th_{q_*}$ is fixed in \eq{cestp}). Using the values of other dimension-full parameters in \eq{fnl0} we obtain 
\be\label{fnl1}
f_{NL}\simeq \, 1.4\times 10^{4}.
\ee
This is a very large amount of non-gaussianity that is solely produced in reheating and it is obviously  inconsistent with observations. On the other hand, by comparing \eq{e2} and \eq{fnl1} we observe that although they measure different one loop corrections, the latter has more $\chi$ modes circulating in the loops and it produces a much bigger number. Therefore, the large  amount obtained in \eq{fnl1} can be an artifact of perturbation theory, which might become invalid in this model. It is possible to produce large non-gaussianity in inflationary models (see e.g. \cite{fnl1}), but the single scalar field models generically give $f_{NL}={\cal O}(\e)$, where $\e$ is the slow roll parameter. Although we are not capable of making non-perturbative estimates, our computations show that a large non-gaussianity can be produced during reheating. 

Using a different approach, namely by looking at local nonlinear terms in field  equations generated through interactions, it has also been shown in \cite{yeni2,yeni3,yeni4,yeni5} that parametric resonance effects might generate large non-gaussianity. Specifically, in \cite{yeni5} the chaotic $\l  \f^4$ model is considered and it is found that for a certain range of parameters one has $f_{NL}>{\cal O}(1000)$. As long as the parametric resonance effects are taken into account, $\l\f^4$ and $m^2\f^2$ models are very similar to each other and thus our result \eq{fnl1} perfectly agrees with \cite{yeni5}. 

\subsection{The tensor power spectrum} 

\begin{figure}
\centerline{
\includegraphics[width=6cm]{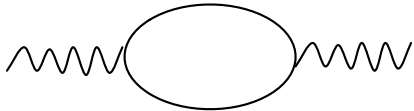}}
\caption{The 1-loop graph arising from the interaction Hamiltonian \eq{h3} that contributes to the graviton two point function $\lf\cc_{ij}\cc_{kl}\rg$  during reheating.}
\label{fig3}
\end{figure}

The interaction Hamiltonian \eq{h3} also modifies the tensor power spectrum due to the last term involving the graviton coupling. One may first think that this interaction is suppressed by $1/a^2$, however this factor simply converts the two comoving momenta arising from the two partial derivatives to the physical scale. The tensor field $\cc_{ij}$ is similar to a spectator field since its background value vanishes. As a result, the tensor power spectrum is not affected by the (infinitesimal)  changes of the  spacetime slicing and the gauge can be fixed in a natural way without giving rise to any complications. Moreover, unlike the $\z$ propagator, the tensor propagator does not contain any singularities. The correction corresponding to \eq{h3} can be pictured as in Fig. \ref{fig3}. 

Using \eq{inp} for $\cc_{ij}(t,\vec{x})\cc_{kl}(t,\vec{y})$ with $N=2$, which gives the first nonzero contribution, and applying the identity $[AB,C]=A[B,C]+[A,C]B$, one finds terms with single or two graviton commutators. It is easy to see that the terms with two graviton commutators are suppressed by $1/a^3$ and hence they become completely negligible. A straightforward calculation then gives the following one loop correction to the tensor power spectrum in momentum space 
\be\label{ccl}
P_k^\cc(t_F)^{(1)}\simeq P_k^{\cc(0)}\fr{4i}{9M_p^2}\int_{t_R}^{t_F}dt_2\int_{t_R}^{t_2}dt_1\, a(t_1)\,a(t_2)\,\left[g(t_2)-g(t)\right]\left[H(t_1,t_2,k)-c.c\right],
\ee
where $g(t)$ is defined in \eq{fg} and 
\be\label{th}
H(t_1,t_2,k)=\fr{1}{(2\pi)^3}\int d^3q \,q^2(k+q)^2\, \chi_{q}(t_1)\chi_{k+q}(t_1)\chi_{q}^*(t_2)\chi_{k+q}^*(t_2).
\ee
In \eq{ccl} we reintroduce the Planck mass $M_p$, which can be fixed either by dimensional analysis or by keeping track of its presence starting from the action \eq{a}. Once again, the one loop correction in momentum space becomes a multiple of the tree level power spectrum. This is mainly because of the fact that the expectation value $\lf [O_{ij}(t_2,\vec{z}_2), O_{kl}(t_1,\vec{z}_1)]\rg$, which appears due to last term of the interaction Hamiltonian \eq{h3}, produces $\d_{ik}\d_{jl}+\d_{il}\d_{kj}$ and this index structure acting on the polarization tensor $\Pi_{ijkl}$, which is  introduced in \eq{pt}, gives the same tensor. 

Converting the comoving integration variable in \eq{th} to the physical scale generates the power $a^7$, and this factor together with $a(t_1)a(t_2)$ in \eq{ccl} completely compensate the suppressions of the mode functions $\chi_q$ and the $1/a^3$ decay of the function $g(t)$. As before, the change in the spectral index is negligible due to the large hierarchy between the superhorizon scale $k$ and the scale $q_*$ characterizing the instability band. Therefore, in \eq{ccl} one may ignore the $k$ dependence, set $q=q_*$ and let $d^3q\to 4\pi q_*^3$.  For the $\chi$ modes, one may use \eq{c-amp} and \eq{cestp}. Finally, to estimate the time integral, we introduce the time dependence of the background quantities using \eq{fh}. As a result we find 
\be\label{corten}
P_k^\cc(t_F)^{(1)}\simeq P_k^{\cc(0)} \fr{8}{27\pi^2M_p^2}\left[\fr{C_3}{m^2} \right]\left[\fr{1}{H(t_F)}\right]\left[q_*^7 |\chi_{q_*}|^4\right],
\ee
where 
\be\label{c3}
C_3=\int_{mt_F-2\pi}^{mt_F}mdt_2\int_{mt_F-2\pi}^{mt_2}mdt_1\fr{t_F^{8/3}}{(t_1t_2)^{4/3}}\left[\fr{t_F}{t_2}-1\right]\sin[2\th_{q_*}(t_2)-2\th_{q_*}(t_1)].
\ee
For our canonical set of parameters \eq{s1}, we numerically integrate \eq{c3} that yields $C_3\simeq 0.029$. Using \eq{s2} for the Hubble parameter one finds 
\be\label{e3}
P_k^\cc(t_F)^{(1)}\simeq 5\times10^{-5} P_k^{\cc(0)}. 
\ee
The reason for this correction to be small compared to the scalar power spectrum \eq{e2} is that  the factor $g^4\Phi^4$ in \eq{e1} is replaced by $q_*^4$ in \eq{corten} due to different forms of interactions in \eq{h3}, and one has $g\Phi\simeq 7q_*$. Nevertheless, the modification \eq{e3} is much larger than the quantum corrections that arise during inflation, which are suppressed by the ratio $H/M_p$ \cite{mal}. 

\section{Some higher order interactions and loops} \label{sec3}

The results of the previous section show that  cubic interactions involving two $\chi$ fields modify the scalar and the tensor power spectra and give rise to non-gaussianity. Although the interaction Hamiltonian \eq{h3} is cubic, the first nonzero contributions come from \eq{inp} with $N=2$ for the scalar and the tensor power spectra, and with $N=3$ for the three point function. The corresponding one loop corrections are sixth and ninth order in fluctuations, respectively. 

In this section, we consider some higher order (e.g. fourth and fifth order) interactions, again involving two $\chi$ fields, and calculate the corresponding one loop effects. Our aim in considering such interactions is two fold. First, we would like to use \eq{inp} with $N=1$. Therefore, by a naive counting in perturbation theory the effects are supposed to be more prominent than the ones we have studied in the previous section (although this turns out to be incorrect as we will see below). Second, the loop effects calculated in the previous section involve the commutators of the $\chi$ fields and thus one must carefully treat the phase factors as we did in \eq{f5}. The loop corrections we consider in this section demonstrate the modifications more directly. 

\subsection{The scalar power spectrum and non-gaussianity}

Starting from the action \eq{a}, one may obtain the following terms in the interaction Hamiltonian 
\be\label{h4}
H_I=\int d^3 x\, a^3\, e^{3\z}\rho_\chi+...=\fr92 \int d^3 x\, a^3\, \left[ \z^2+\z^3\right]\rho_\chi + ...
\ee
where $\rho_\chi$ is the energy density of the $\chi$ field given by 
\be
\rho_\chi=\fr12 g^2\f^2\chi^2+\fr{1}{2a^2}(\del\chi)^2+\fr{1}{2}\dot{\chi}^2. 
\ee 
The first term in \eq{h4} contributes to the scalar power spectrum and the second one produces scalar non-gaussianity. Note that the linear $\z$ term in \eq{h4} agrees with the cubic hamiltonian in \eq{h3}. 

\begin{figure}
\centerline{
\includegraphics[width=5cm]{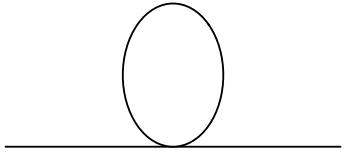}}
\caption{The 1-loop graph arising from the interaction Hamiltonian \eq{h4} that contributes to the scalar power spectrum during reheating.}
\label{fig4}
\end{figure}

Let us first consider the one loop correction to the scalar power spectrum arising from \eq{h4} that can be pictured as in Fig. \ref{fig4}. Using \eq{inp} for the $\z(t,\vec{x})\z(t,\vec{y})$ with $N=1$,  a straightforward calculation gives
\be\label{e4}
P_k^\z(t_F)^{(1)}\simeq 18 P_k^{\z(0)}\int_{t_R}^{t_F} dt_1 a(t_1)^3\lf\rho_{\chi}(t_1)\rg \left[f(t_1)-f(t_F)\right].
\ee
This equation clearly shows how the correction enlarges in time during preheating as the energy density $\lf\rho_{\chi}\rg$ increases as a result of  $\chi$ particle creation. Note that \eq{e4} only modifies the amplitude of the spectrum since the correction multiplying the tree level result does not depend on the external momentum $k$. At the end of the first stage of preheating the energy density of the created $\chi$ particles catches up the background energy density, which gives $\lf\rho_{\chi}(t_1)\rg\simeq 3H^2M_p^2$. Reading  $f(t)$ from \eq{fs}, it is easy to see that 
\be\label{e44}
P_k^\z(t_F)^{(1)}\simeq {\cal O}\left(\fr{H(t_F)}{m}\right) P_k^{\z(0)}.
\ee
Indeed, using \eq{fh} for the background quantities one finds that 
\be
P_k^\z(t_F)^{(1)}\simeq  \fr{12H(t_F)}{m}\int_{mt_F-2\pi}^{mt_F} mdt\left[\fr{t_F}{t}-1\right] P_k^{\z(0)}\simeq 0.05  P_k^{\z(0)},
\ee
where, as before,  we restrict the time integral to the last inflaton oscillation cycle. 

One may find other terms in the interaction Hamiltonian that modifies the scalar power spectrum. For instance, by introducing $\exp(3\z)$ factor in \eq{h3}, which arises from $\sqrt{h}$,  one obtains a fourth order  term
\be\label{h44}
H^{(4)}_I=3 \int d^3x \,a^3\,\z\dot{\z}O_1. 
\ee
After using \eq{h44} in \eq{inp} with $N=1$,  one encounters terms either with $\lf \z\z\rg[\dot{\z},\z]$ or $\lf \z\dot{\z}\rg[\z,\z]$. It is easy to see that the latter is suppressed by $1/a^3$ and the former yields 
\be\label{e5}
P_k^\z(t_F)^{(1)}\simeq 6 P_k^{\z(0)}\int_{t_R}^{t_F} dt_1 a(t_1)^3\lf O_1(t_1)\rg \fr{df}{dt}(t_1).
\ee
From \eq{o1}, one has $\lf O_1\rg\simeq\lf\rho_\chi\rg/H$ and using \eq{fgn} we obtain 
\be\label{e55}
P_k^\z(t_F)^{(1)}\simeq {\cal O}\left(\fr{H(t_F)}{m}\right)  P_k^{\z(0)}.
\ee
The main conclusion here is that although the corrections \eq{e44} and \eq{e55} correspond to lower order in perturbation theory, they give smaller contributions compared to \eq{e2}.  

\begin{figure}
\centerline{
\includegraphics[width=3cm]{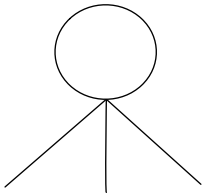}}
\caption{The 1-loop graph arising from the interaction Hamiltonian \eq{h4} that contributes to the three point function $\lf\z\z\z\rg$.}
\label{fig5}
\end{figure}

The fifth order term $\z^3\rho_\chi$ in \eq{h4} corrects the three point function and thus it gives rise to non-gaussianity. The corresponding graph is pictured in Fig. \ref{fig5}. Using \eq{inp} for $\z(t,\vec{x})\z(t,\vec{y})\z(t,\vec{z})$ with $N=1$ and using the definition of the three point function in momentum space given in \eq{def3}, one  finds that 
\be\label{tii} 
P(k_1,k_2,k_3)^{(1)}\simeq \fr{27}{(2\pi)^6}
\int_{t_R}^{t_F} dt_1\,a(t_1)^3 \lf\rho_{\chi}(t_1)\rg \left[f(t_1)-f(t_F)\right]\,P_{k_1}^{\z(0)}P_{k_2}^{\z(0)}+\textrm{cyclic}.
\ee
From \eq{60},  the corresponding $f_{NL}$ parameter can be calculated as
\be
f_{NL}\simeq  45 \int_{t_R}^{t_F} dt_1\,a(t_1)^3 \lf\rho_{\chi}(t_1)\rg \left[f(t_1)-f(t_F)\right].
\ee
As in \eq{h44}, by introducing $\exp(3\z)$ factor in \eq{h3} gives the following interaction Hamiltonian: 
\be\label{h444}
H^{(4)}_I=\fr{9}{2} \int d^3x \,a^3\,\z^2\dot{\z}O_1.
\ee
It's contribution  to $f_{NL}$ can be found as  
\be
f_{NL}\simeq  15 \int_{t_R}^{t_F} dt_1\,a(t_1)^3 \lf O_1(t_1)\rg \fr{df}{dt}(t_1).
\ee
In both of these cases it is easy to estimate the integrals so that  
\be\label{fc} 
f_{NL}\simeq {\cal O}\left(\fr{H(t_F)}{m}\right). 
\ee
Therefore a small amount of non-gaussianity is produced by these interactions. As in the case of the power spectrum, the loop corrections to $f_{NL}$ coming from the interactions that can be pictured as in Fig. \ref{fig5} become much smaller than the previous one \eq{fnl1}. 

\subsection{Fourth order interactions that has the form $\cc\cc\chi\chi$ and the tensor power spectrum} 

Till now in this section we have considered some higher order interactions that modify the scalar power spectrum and the $f_{NL}$ parameter. It is clear that in a systematic study  one should work out the complete fourth order action to determine the corrections more accurately. In that case, the lapse $N$ and the shift $N^i$ must be solved up to second order. This is a complicated calculation and the complete fourth order action is not very illuminating for the scalar field. However, the interactions studied above are generic enough to indicate that other corrections to the scalar power spectrum and $f_{NL}$ will be similar to the ones  found above. 

In this subsection, we determine the complete fourth order action involving the interactions of the tensor field $\cc_{ij}$ and the reheating scalar $\chi$. Our aim is again to compare the corresponding corrections with \eq{e3} to see how the perturbation theory is working. Since we solely  concentrate on the tensor modes we set 
\be
\z=0.
\ee
(recall that we have been working in the $\vf=0$ gauge). The quartic interactions involving $\cc_{ij}$ and $\chi$ are necessarily in the from $\cc\cc\chi\chi$ since the background values of $\cc_{ij}$ and $\chi$ are zero. Similarly, there is no linear term in $N$ and $N^i$ after one sets $\z=0$. We define
\bea
&&N=1+N^{(2)},\\
&&N^i=N^{(2)i}_T+\del_i\psi^{(2)},
\eea
where $\del_iN^{(2)i}_T=0$. To determine these second order quantities, one may use the exact solution for the lapse 
\be
N=\sqrt{\fr{B}{A}},
\ee
where $A$ and $B$ are defined in \eq{aa} and \eq{bb}, and work out the momentum constraint, which reads
\be
D_i\left(\fr{1}{N}\left[K^i{}_j-\d^i_j K\right]\right)=\fr{1}{N}\left(\dot{\chi}-N^i\del_i\chi\right)\del_j\chi,
\ee
where $K_{ij}$ and $K$ are defined above \eq{aa}. Up to second order in fluctuations, the Ricci scalar $R^{(3)}$ of the constant time hypersurface can be found as 
\be
R^{(3)}=-\fr{1}{4a^2}(\del_i\cc_{jk})(\del_i\cc_{jk}).
\ee
After a relatively long but straightforward calculation we find
\bea
&&\del^2N^{(2)}=\fr{1}{8H}\del_j(\dot{\cc}_{ik}\del_j\cc_{ik})+\fr{1}{2H}\del_j(\dot{\chi}\del_j\chi),\label{lsf}\\
&&\del^2\psi^{(2)}=-\fr{1}{16H}\dot{\cc}_{ij}\dot{\cc}_{ij}-\fr{1}{16Ha^2}(\del_i\cc_{jk})(\del_i\cc_{jk})-\fr{1}{4H}\dot{\chi}^2-\fr{1}{4Ha^2}(\del_i\chi)(\del_i\chi)-\fr{g^2\f^2}{4H}\chi^2-\fr{m^2\f^2}{2H}N^{(2)}.\nn
\eea
Similarly, the transverse part of the shift reads
\be
\del^2N^{(2)i}_T=\fr12\del_i\fr{1}{\del^2}\del_k(\dot{\cc}_{mn}\del_k\cc_{mn})-\fr12 \cc_{jk}\del_j\dot{\cc}_{ki}+\fr12\dot{\cc}_{jk}\del_j\cc_{ki}-\fr12\dot{\cc}_{jk}\del_i\cc_{jk}-2\dot{\chi}\del_i\chi+\fr{2}{\del^2}\del_i(\del_j(\dot{\chi}\del_j\chi)).\label{sf}
\ee
In all these expressions the indices are contracted with the Kronecker delta and we set $M_p=1$. 

Before discussing the loop corrections, it is interesting to check the validity of the perturbation theory from the quadratic expressions given for the lapse and the shift. As discussed in \cite{coh2}, the perturbation theory is applicable if one has    
\be\label{vp}
\lf N^{(2)}\rg\ll1,\hs{10}\lf\del_i N^i\rg=\lf\del^2\psi^{(2)}\rg\ll H.
\ee
While the first condition is needed for keeping the time coordinate to be proper, the second ensures that the original foliation of the spacetime that is presumed for perturbation theory is not destroyed by the fluctuations. It is obvious that the terms containing the $\chi$ field are dangerous for the conditions \eq{vp}. From \eq{lsf} we find 
\be
\lf N^{(2)}\rg\simeq \fr{\o_{q_*}}{2HM_p^2}\lf\chi^2\rg.
\ee
Since near the end of the first stage $\lf\chi^2\rg\simeq m^2/g^2$ and $\o_{q_*}\simeq g\Phi$, one has
\be
\lf N^{(2)}\rg\simeq \fr{m^2\Phi}{gHM_p^2}.
\ee
From this expression it is easy to see that the first condition in \eq{vp} is safe. On the other hand, using \eq{lsf} the second condition in \eq{vp} demands  
\be
\fr{\lf\rho_\chi\rg}{H^2M_p^2}\ll1.
\ee
It is clear that when the energy density of $\chi$ particles catches up the background energy density, i.e. $\lf\rho_\chi\rg\simeq H^2M_p^2$, and this condition is invalidated. This result is independent of our loop considerations and separately indicates  the failure of the perturbation theory in this model. 

\begin{figure}
\centerline{
\includegraphics[width=4.4cm]{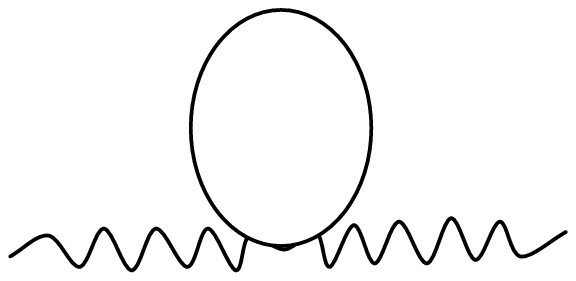}}
\caption{The 1-loop graph arising from the action  \eq{s4} that contributes to the tensor spectrum $\lf\cc\cc\rg$.}
\label{fig6}
\end{figure}

Returning to the interactions involving $\chi$ and $\cc_{ij}$, one can use the solutions for the lapse and the shift in \eq{a} to find  
\be\label{s4}
S^{(4)}_{\cc\cc\chi\chi}=\fr12\int a^3\left[-\fr{1}{2a^2}\cc_{ik}\cc_{jk}\del_i\chi\del_j\chi-2N^{i}_{\cc}\dot{\chi}\del_i\chi+(2m^2\f^2)N_{\cc}N_{\chi}+(K_{ij}K^{ij}-K^2)^{(4)}\right],
\ee
where a subindex on $N$ or $N^i$ indicates that only  the relevant  terms must be kept in \eq{lsf} and \eq{sf}. To fix the action completely, one should also determine the fourth order terms in $K_{ij}K^{ij}-K^2$, however we will not need them for  our analysis below.   The corresponding corrections can be pictured as in Fig. \ref{fig6}. 

It is clear that in \eq{s4} the terms containing $\del_i\cc_{jk}$ are suppressed by the factors $\hat{k}/M_p$, $\hat{k}/q_*$ or $\hat{k}/H$, where $\hat{k}=k/a$ is the physical superhorizon scale of interest. Similarly, since a time derivative acting on $\cc_{ij}$ kills the constant piece in the mode function, the terms containing $\dot{\cc}\dot{\cc}$ are completely negligible because they decay like $1/a^3$. Likewise, the terms that has the structure $\dot{\cc}\cc$ would be equivalent to $H\cc\cc$ (note that these appear from the commutator $[\cc,\dot{\cc}]$).  As a result, we conclude that the first term in \eq{s4} gives the typical correction to the tensor power spectrum and the corresponding interaction Hamiltonian becomes
\be\label{hc4}
H_I^{(4)}=\fr{a}{4}\int \cc_{ik}\cc_{jk}\del_i\chi\del_j\chi.
\ee
A straightforward calculation gives the following one loop correction to the tensor power spectrum:
\be
P_k^\cc(t_F)^{(1)}\simeq P_k^{\cc(0)}\fr83 \int_{t_R}^{t_F}dt\, a(t)\, \lf \del_i\chi(t)\del_i\chi(t) \rg \, [g(t)-g(t_F)].
\ee
To estimate this correction, we first note that $ \lf \del_i\chi\del_i\chi \rg\simeq a^2q_*^2\lf\chi^2\rg$. We then focus on the last inflaton oscillation cycle in which $\lf\chi^2\rg$ reaches its maximum value. Treating $\lf\chi^2\rg$ as a constant and using \eq{fh} for the background quantities one may estimate  
\be
P_k^\cc(t_F)^{(1)}\simeq P_k^{\cc(0)}\fr{16q_*^2\lf\chi(t_F)^2\rg}{9mH(t_F)M_p^2}\int_{mt_F-2\pi}^{mt_F}mdt\,\,\fr{t^2}{t_F^2}\left[\fr{t_F}{t}-1\right].
\ee
For our canonical case \eq{s1}, the integral can be evaluated numerically to yield $0.28$. Using \eq{ce} and the values of the other dimension-full parameters we obtain
\be\label{e42} 
P_k^\cc(t_F)^{(1)}\simeq  10^{-6}P_k^{\cc(0)}
\ee
We see that this correction is two orders of magnitude smaller than \eq{e3}. As before, a correction which is supposed to be larger according to the naive counting in perturbation theory turns out to be smaller. Note that both corrections \eq{e3} and \eq{e42} are still larger than the quantum effects produced during inflation, which are characterized by the ratio $H/M_p\simeq 10^{-8}$ \cite{mal}. 

\section{Conclusions} \label{sec4}

In a recent work \cite{ali1}, one of us has shown that the loop quantum effects during reheating significantly modify the scalar power spectrum. In this paper, in an attempt to extend the findings of \cite{ali1} we consider how loops in reheating produce non-gaussianity and affect the tensor power spectrum in the chaotic $m^2\f^2$ model. Based on the tree level results, this model is actually ruled out by 95\% confidence level by the Planck data (provided the running of the index is neglected), however our findings show that quantum effects during reheating can change this conclusion since the corresponding corrections can alter the tree level results appreciably. 

In most of the scalar field inflationary models, inflation is followed by a period of coherent inflaton oscillations where the background is still homogeneous and isotropic. This phase continues until the backreaction effects are set in. As pointed out  in \cite{ca}, in such a background causality does not preclude the emergence of  the superhorizon effects because by coherency the same physical influence can appear at different positions at the same time.  Therefore,  the quantum effects can be important for cosmological variables in the first stage of reheating. On the other hand, it is known that the entropy perturbations can cause nontrivial superhorizon evolution of the curvature perturbation. Consequently, it is not surprising to see that the effects of entropy modes circulating in the loops become significant, especially  in the parametric resonance regime. Indeed, we observe that  the corrections get larger as the number of $\chi$ modes circulating in the loops increases, which indicates that the perturbation theory might become invalid. 

It is well known that in the chaotic model we have studied, the curvature perturbation $\z$ becomes an ill defined variable during reheating. Because of that reason in \cite{ali1}, the calculations have been carried out in the $\z=0$ gauge till the end of the first stage of reheating and then a gauge transformation has been applied to read the $\lf\z\z\rg$ correlation function. In this paper, we utilize a different strategy and smooth out the spikes of $\z$ by using the averaged out background variables in the quadratic $\z$-action. As it is shown above, the results obtained in this way is consistent with \cite{ali1} and thus our conclusions about the scalar power spectrum (and non-gaussianity) are firm. Note that the tensor calculation is free from the gauge fixing issues. 

It is possible to develop the results of this paper in different directions. Due to the importance of the chaotic $m^2\f^2$ model,  it would be valuable to perform a full numerical check  of the loop corrections that are estimated in this paper. It would also be crucial to see whether the loops in reheating modify the predictions of the models that are favored by Planck data, like the Starobinsky model \cite{strm}.  Finally, it would be interesting to determine the loop effects when the inflaton decay occurs perturbatively. In that case while the reheating scalar modes cannot take large values, the decay process is completed in a long time that might enhance the quantum effects, since according to in-in formalism \eq{inp}, the quantum corrections are proportional to the duration of the process.  

\begin{acknowledgments}
N. Kat\i rc\i  \ thanks Bo\~{g}azi\c{c}i University for the financial support provided by the Scientific Research Fund with project no: 7128
\end{acknowledgments}

\end{document}